\documentclass[twocolumn, twocolappendix]{aastex63}
	
\usepackage[encapsulated]{CJK}
\usepackage{ucs}
\usepackage[utf8x]{inputenc}
\newcommand{\cntext}[1]{\begin{CJK}{UTF8}{bsmi}#1\end{CJK}}

\usepackage[italicdiff]{physics} 
\usepackage{amsmath}

\usepackage{graphicx}

\usepackage{braket}

\newcommand*{\ve}[1]{\boldsymbol{#1}}
\newcommand{\m}[1]{\underline{\underline{#1}}}
\newcommand{\wt}{\widetilde}
\renewcommand{\ol}{\overline}

\shorttitle{\textit{A priori} test for astrophysical turbulence}
\shortauthors{Hu \& Chiang}

\begin{document}


\title{\textit{A priori} Validation of Subgrid-scale Models for Astrophysical Turbulence}
\author[0000-0002-9235-3529]{Chia-Yu Hu (\cntext{胡家瑜}) }
\affiliation{Center for Computational Astrophysics, Flatiron Institute, 162 5th Ave, New York, NY 10010, USA}
\affiliation{Max-Planck-Institut f\"{u}r Extraterrestrische Physik, Giessenbachstrasse 1,
D-85748 Garching, Germany}
\author[0000-0003-2043-4618]{Chi-Ting Chiang (\cntext{蔣季庭}) }
\affiliation{Physics Department, Brookhaven National Laboratory, Upton, NY 11973, USA}

\begin{abstract}

We perform \textit{a priori} validation tests of subgrid-scale (SGS) models for the 
turbulent transport of momentum, energy and passive scalars.
To this end,
we conduct two sets of high-resolution hydrodynamical simulations with a Lagrangian code:
an isothermal turbulent box with rms Mach numbers of 0.3, 2 and 8,
and the classical wind tunnel where a cold cloud traveling through a hot medium gradually dissolves due to fluid instabilities.
Two SGS models are examined: the eddy diffusivity (ED) model widely adopted in astrophysical simulations and the ``gradient model'' due to \citet{1979JFM....91....1C}. 
We find that both models predict the magnitude of the SGS terms equally well (correlation coefficient $>0.8$).
However,
the gradient model provides excellent predictions for the orientation and shape of the SGS terms
while the ED model predicts both poorly,
indicating that isotropic diffusion is a poor approximation to the instantaneous turbulent transport.
The best-fit coefficient of the gradient model is in the range of [0.16, 0.21] for the momentum transport,
and the turbulent Schmidt number and Prandtl number are both close to unity, in the range of [0.92, 1.15].


\end{abstract}

\section{Introduction} \label{sec:intro}

Turbulence is ubiquitous in astrophysics,
as the length scale of interest is usually orders of magnitude larger than the dissipation length scale.
This inevitably implies a huge dynamical range which poses a formidable challenge to follow the full turbulence cascade from the largest scale to the smallest in numerical simulations.
As such,
one has to adopt subgrid-scale (SGS) models to account for the unresolved turbulent transport.

In particular,
in Lagrangian simulations where cells\footnote{We will use the terms ``cell'' and ``particle'' interchangeably.} are advected with the flows,
there is by construction no mass exchange between cells.
Therefore,
for any scalar field (e.g. metallicity, molecular species, dust, etc.) passively advecting with the flows,
there is no SGS transport or mixing mechanism even if neighboring cells have very different scalar concentrations.
This can cause issues when important physical processes are operating at the resolution scale.
In galaxy formation,
for example,
radiative cooling and star formation typically operate on a cell-by-cell basis,
and metal enrichment also takes place locally among a few cells.
The lack of SGS mixing mechanism implies that once a gas cell is enriched, 
its metallicity will remain high with no ways of dropping in the future,
which leads to spurious phenomena such as excessive cooling or stars forming with extremely high metallicities.

As a consequence,
it is common to include explicit SGS models for scalar transport in Lagrangian simulations,
usually by adding a scalar diffusion equation with the diffusivity determined by local flow properties.
\citet{2009MNRAS.392.1381G} determined the diffusivity by the velocity dispersion based on dimensional analysis.
\citet{2010MNRAS.407.1581S} followed the eddy diffusivity (ED) model \citep{1963MWRv...91...99S} where the diffusivity is determined by the magnitude of the velocity shear multiplied by a constant coefficient.
This model has since been widely applied in Lagrangian simulations,
(e.g. \citealp{2012ApJ...760...50S, 2013MNRAS.434.3142A, 2014MNRAS.445..581H, 2015MNRAS.454...83W, 2016ApJ...824...57C, 2017AJ....153...85S, 2017MNRAS.471..144S, 2018MNRAS.475.3283C, 2019MNRAS.483.3363H, 2019ApJ...885...33H}).
\citet{2017MNRAS.467.2421C} quantified the diffusivity using ensemble-averaged results from high-resolution turbulent box simulations and concluded that the coefficient cannot be constant.
\citet{2019MNRAS.483.3810R} adopted the dynamical ED model \citep{1991PhFlA...3.1760G} allowing a spatially variable coefficient calculated with an extra filter.
The common assumption of all these models is that turbulent transport can be approximated by an isotropic diffusion process.
However,
a rigorous validation test, 
where results of high-resolution simulations are coarse-grained to verify the SGS models\footnote{It is often referred to as the ``\textit{a priori} test'' in the literature of turbulence modeling (see, e.g., reviews by \citealp{1999PrAeS..35..335P} and \citealp{2000AnRFM..32....1M}).},
has never been done.

On the other hand,
in Eulerian simulations,
scalar diffusion is usually not explicitly taken into account.
The rationale behind this is that numerical diffusion,
which originates from truncation errors and operates at the resolution scale,
serves as an SGS model of turbulent dissipation.
By the same token,
turbulent transports of momentum and energy are also often ignored in both Eulerian and Lagrangian simulations.
However,
the exact mechanism of numerical dissipation is intractable and obviously solver-dependent 
(e.g., Eulerian vs. Lagrangian, low-order vs. high-order methods, different Riemann solvers, etc.).
A more robust approach would require explicit SGS models whose effects dominate over numerical diffusion.
\citet{2011A&A...528A.106S} proposed and tested an SGS model for momentum transport with Eulerian codes.
Their model was later on applied in cosmological simulations \citep{2014MNRAS.440.3051S}
and in idealized galaxy-scale simulations \citep{2016ApJ...826..200S}.
With a Lagrangian code,
\citet{2019MNRAS.483.3810R} applied their dynamical ED model to momentum and energy diffusion (in addition to scalar diffusion) in both idealized test cases and more realistic simulations of galaxy formation.


In the literature of terrestrial turbulence modeling,
there is another type of SGS model called the ``gradient model'' first proposed by \citet{1979JFM....91....1C}.
This model takes a more agnostic view on the physics of SGS transport and is based purely on Taylor expansion of the SGS terms.
It has shown to be an improvement over the ED model in several cases (see \citealp{2000AnRFM..32....1M} and references therein).
However, 
its application in astrophysical problems remains rare so far.

In this paper,
we perform \textit{a priori} tests to verify SGS models for momentum, energy and passive scalars.
To this end, 
we conduct two sets of hydrodynamical simulations:
one is an idealized isothermal turbulent box in both subsonic and supersonic regimes,
and the other is the classical wind tunnel problem
where a cold and dense cloud gradually dissolves in a hot and diffuse medium due to fluid instabilities.
We will show that all ED-type models predict SGS terms whose orientations and shapes
correlate poorly with the true SGS terms.
On the other hand,
the gradient model
provides much accurate predictions in both steady-state and transient problems.
Our \textit{a priori} analysis code is publicly available\footnote{https://github.com/huchiayu/Lapriori.jl}.

This paper is organized as follows.
In Sec. \ref{sec:theory},
we outline the theoretical background of turbulent SGS terms and describe the two models to be tested.
In Sec. \ref{sec:results},
we present our two sets of numerical simulations and the results of our \textit{a priori} analysis.
In Sec. \ref{sec:sum},
We discuss the potential caveats of adopting SGS models and summarize our work.

\section{Theoretical Formulation}\label{sec:theory}

In this section, 
we give an overview of the theoretical formulation for the SGS models.
Vectors and tensors are expressed either in index notation with the Einstein summation convention
or in bold (for vectors) and double-underbar symbols (for tensors).

\subsection{Coarse-grained Fluid Equations}

The fluid equations are composed of the continuity equation,
\begin{equation}
\pdv{\rho}{t} + \pdv{( \rho v_j)}{x_j} = 0,
\end{equation}
the momentum equation,
\begin{equation}
\pdv{(\rho v_i)}{t} + \pdv{( \rho v_i v_j + P \delta_{ij} ) }{x_j}= 0,
\end{equation}
and the energy equation,
\begin{equation}
\pdv{(\rho e)}{t} + \pdv{\big( (\rho e + P) v_j \big)}{x_j} = 0,
\end{equation}
where $\rho$, $v_j$ and $P$ are the density, velocity and pressure of the fluid, respectively.
The specific total energy is $e = u + v^2/2$ where $u$ is the specific thermal energy and $v^2 = v_k v_k$.
In addition,
the transport equation of a passive scalar $\phi$ is
\begin{equation}
\pdv{(\rho\phi)}{t} + \pdv{( \rho \phi v_j)}{x_j} = 0.
\end{equation}

In practical simulations,
the cell size defines a minimal resolvable length scale.
For any given field function $f(\ve{x})$,
we can convolve it with a kernel function $W(\ve{x} - \ve{x}',h)$ and obtain a coarse-grained field:
\begin{equation}\label{eq:filter}
\ol{f}(\ve{x}) = \int f(\ve{x'}) W(\ve{x} - \ve{x'},h) d^3x 
\end{equation}
where $h$ is the filter width corresponding to the cell size.
The convolution averages out small-scale information below $h$ and 
it is $\ol{f}$ (rather than $f$) that is actually present in simulations.

The coarse-grained continuity equation follows
\begin{equation}
\pdv{\ol{\rho}}{t} + \pdv{x_j} ( \ol{\rho} \wt{v_j}) = 0, 
\end{equation}
where $\wt{f} \equiv {\ol{\rho f}} / {\ol{\rho}}$ is the ``Favre filter''
commonly used  for compressible fluid such that there are no unclosed terms in the continuity equation.
The coarse-grained momentum equation,
\begin{equation}
\pdv{(\ol{\rho} \wt{v_i} )}{t} + \frac{\partial }{\partial x_j} 
( \ol{\rho} \wt{v_i}\wt{v_j} + \ol{P}\delta_{ij} + \ol{\rho} \tau_{ij}) = 0, \\
\end{equation}
contains an unclosed term referred to as the SGS stress tensor
\begin{equation}
\tau_{ij} \equiv  \wt{v_i v_j} -  \wt{v_i}\wt{v_j}. \label{eq:true_tau}
\end{equation}
The coarse-grained energy equation follows
\begin{equation}
\pdv{(\ol{\rho} \wt{e})}{t} + \pdv{x_j}\big( (\ol{\rho} \wt{e} + \ol{P}) \wt{v_j} + \ol{\rho}( \theta_j + d_j ) + p_j \big) = 0,
\end{equation}
where $\theta_j$, $d_j$ and $p_j$ are the SGS heat flux, kinetic energy flux and pressure flux, respectively, viz.,
\begin{align}
\theta_{j} &\equiv  \wt{u v_j} -  \wt{u}\wt{v_j}, \label{eq:true_qe} \\
d_j &\equiv  \wt{v_k v_k v_j} -  \wt{v_k v_k}\wt{v_j}, \\
p_j &\equiv  \ol{P v_j} -  \ol{P}\wt{v_j}.
\end{align}
The SGS kinetic energy flux can be approximated as $d_j \approx \tau_{jk} \wt{v_k}$ \citep{1999PrAeS..35..335P},
while the SGS pressure flux can be further expressed as $p_j = (\gamma - 1) \ol{\rho} \theta_j$ 
where we have used the coarse-grained equation of state 
\begin{equation}
\ol{P} = \ol{(\gamma - 1)\rho u} = (\gamma - 1)\ol{\rho} \wt{u}.
\end{equation}
Finally, the coarse-grained transport equation for passive scalar is
\begin{equation}
\pdv{(\ol{\rho}\wt{\phi})}{t} + \pdv{x_j} ( \ol{\rho} \wt{\phi}\wt{v_j} + \ol{\rho}q_j) = 0, 
\end{equation}
where 
\begin{equation}
q_j \equiv  \wt{\phi v_j} -  \wt{\phi}\wt{v_j},  \label{eq:true_qphi}
\end{equation}
is the SGS scalar flux.

The SGS terms, $\tau_{ij}$, $\theta_j$ and $q_j$, are extra unknowns introduced by filtering and thus they must be modeled (viz., the ``closure problem'').
They represent the interaction between the resolved and subgrid scales.


\subsection{Eddy Diffusivity (ED) Model}

The ED model is the most widely used SGS model.
If we define the anisotropic part of the SGS stress tensor by subtracting its trace, viz.,
\begin{equation}
\tau^*_{ij} \equiv \tau_{ij} - \frac{\tau_{kk}}{3} \delta_{ij},
\end{equation}
the ED model follows\footnote{We use the overhat notation $\hat{\tau}$ to emphasize that this is a model prediction as opposed to the true SGS stress tensor $\tau$. The same applies to the other SGS terms.}
\begin{equation}
\hat{\tau}^*_{ij} = - a_\tau^{\rm ED} h^2 | \m{\wt{S}} | \wt{S}_{ij} \label{eq:eddy_tau}
\end{equation}
where  $a_\tau^{\rm ED}$ is a constant coefficient\footnote{The coefficient is sometimes expressed alternatively as $a_\tau^{\rm ED} = \sqrt{2}c_s^2$ where $c_s$ is the so-called ``Smagorinsky constant.'' 
},
\begin{equation}
\wt{S}_{ij} \equiv \frac{1}{2} (\pdv{\wt{v_j}}{x_i} + \pdv{\wt{v_i}}{x_j}) - \frac{1}{3} \pdv{\wt{v_k}}{x_k} \delta_{ij}
\end{equation}
is the (coarse-grained) velocity shear tensor and $| \m{\wt{S}} | \equiv \sqrt{\wt{S}_{ij} \wt{S}_{ij}}$ is the magnitude\footnote{It is also known as the ``Frobenius norm'' of a matrix and is coordinate independent.}
 of the tensor $\wt{S}_{ij}$.
It is important to note that this model only predicts the anisotropic part of $\tau_{ij}$,
which is evident as both $\tau^*_{ij}$ and $\wt{S}_{ij}$ are traceless tensors.
In fact, the trace of $\tau_{ij}$ is related to the specific SGS turbulent kinetic energy, viz.,
\begin{equation}
k\equiv\frac{\tau_{kk}}{2} = \frac{1}{2} (\wt{v_k v_k} -  \wt{v_k}\wt{v_k}).
\end{equation}
For incompressible flows where the energy equation decouples from the momentum and continuity equations,
$k$ is commonly incorporated into the coarse-grained pressure,
and this effective pressure is solved while the thermal pressure is left unknown.
However, the same cannot be done for compressible flows.
Instead,
an additional model for $k$ is required.
For example,
one can augment the fluid equations by introducing a transport equation for $k$ (e.g. \citealp{2011A&A...528A.106S}).
One the other hand,
\citet{2019MNRAS.483.3810R} ignored this term and simply assumed $\tau_{ij} = - a_\tau^{\rm ED} h^2 | \m{\wt{S}} | \wt{S}_{ij}$.
This makes $\tau_{ij}$ traceless by construction and is effectively assuming $k=0$, which is not physically justified.

Similarly,
the models for SGS transport of thermal energy and scalars are, respectively, 
\begin{align}
\hat{\theta}_i   &= -a_\theta^{\rm ED} h^2 | \m{\wt{S}} | \pdv{\wt{u}}{x_i}, \label{eq:eddy_qe} \\
\hat{q}_i &= -a_q^{\rm ED} h^2 | \m{\wt{S}} | \pdv{\wt{\phi}}{x_i}, \label{eq:eddy_qphi}
\end{align}
where $a_\theta^{\rm ED}$ and $a_q^{\rm ED}$ are constant coefficients,
and,
unlike for $\tau_{ij}$, 
there are no missing terms that would require a separate modeling.
The (negative) gradients of $\wt{u}$ and $\wt{\phi}$ determine the orientations of $\theta_i$ and $q_i$, respectively.

As a diffusion equation,
the ED model is always dissipative and thus does not allow for back scattering (turbulent transport from small to large scales).
It has been found to be too dissipative in some cases such as laminar flows.

\subsection{Gradient model}

The gradient model \citep{1979JFM....91....1C} is based on Taylor expansion on the unfiltered fields.
Applying Taylor expansion on $f(\ve{x'})$ around $\ve{x}$, 
the coarse-grained field in Eq. \ref{eq:filter} becomes
\begin{equation}
	\ol{f} = f + \pdv{^2 f}{x_j^2}\frac{M_2 h^2}{6} + \mathcal{O}(h^4),
\end{equation}
where $M_2$ is the second moment of $W$, viz.,
\begin{equation}
	M_2 \equiv \int W(\ve{x} - \ve{x'}) |\ve{x} - \ve{x'}|^2 d^3x ,
\end{equation}
and we have used the fact that the first moment of $W$ vanishes since $W$ is symmetric and isotropic.
It is straightforward to show that the Favre-filtered field follows
\begin{equation} \label{eq:taylorwt}
	\wt{f} = f + \frac{M_2 h^2}{6} \Big(\pdv{^2 f}{x_j^2} + \frac{2}{\rho}\pdv{\rho}{x_k}\pdv{f}{x_k}\Big) + \mathcal{O}(h^4).
\end{equation}
Therefore,
for any functions $f$ and $g$, 
we can write down the following expression:
\begin{equation}
\wt{fg} - \wt{f}\wt{g} = \frac{M_2 h^2}{3} \Big(\pdv{\wt{f}}{x_k}\pdv{\wt{g}}{x_k}\Big) + \mathcal{O}(h^4).
\end{equation}
where we have used the fact that $\partial \wt{f} / \partial x_k = \partial f / \partial x_k + \mathcal{O}(h^2)$ from Eq. \ref{eq:taylorwt}.
For the cubic spline kernel we adopt,
we can obtain analytically that $M_2 = 0.225$.
However,
in practice,
the normalization is left as a free parameter to account for the high-order error.
The gradient model can therefore be summarized as:
\begin{align}
\hat{\tau}_{ij} &= a_\tau^{\rm G} h^2 \pdv{\wt{v_i}}{x_k} \pdv{\wt{v_j}}{x_k}, \label{eq:grad_tau}\\
\hat{\theta}_i   &= a_\theta^{\rm G} h^2 \pdv{\wt{v_i}}{x_k} \pdv{\wt{u}}{x_k},  \label{eq:grad_qe}\\
\hat{q}_i &= a_q^{\rm G} h^2 \pdv{\wt{v_i}}{x_k} \pdv{\wt{\phi}}{x_k},  \label{eq:grad_qphi}
\end{align}
where $a_\tau^{\rm G}$, $a_\theta^{\rm G}$ and $a_q^{\rm G}$ are coefficients that are expected to be close to $M_2/3 = 0.75$ if the high-order error is negligible.
The advantage over the ED model for the $\tau_{ij}$ terms is that the gradient model predicts the entire SGS stress tensor and therefore there is no need for a separate model for $k$.
The gradient model can be viewed as a tensor diffusivity model where the diffusivity is determined by the velocity gradient tensor  ${\partial \wt{v_i}} / {\partial x_k}$,
which contains the information of velocity divergence, shear and vorticity.
In contrast, the diffusivity in the ED model is determined solely by $| \m{\wt{S}} |$.
In addition,
the orientation is not determined solely by the gradients of the transport quantities.
The gradient model is not purely dissipative and it allows for back scattering.

It should be noted that the cubic spline kernel we adopt is, strictly speaking, 
only appropriate for smoothed-particle hydrodynamics (SPH).
For Cartesian mesh codes or adaptive mesh refinement (AMR) codes,
a top-hat kernel is more appropriate.
The top-hat kernel, though not isotropic, is symmetric such that the first moment still vanishes.
For moving mesh codes or meshless finite-volume  methods,
the kernel is no longer symmetric due to the irregular volume partition.\footnote{
For meshless methods,
kernels are distorted due to the normalization by the effective particle number density (i.e. the Shepard correction) 
which makes them asymmetric.}
However,
for these methods,
if we do the Taylor expansion around the centroid rather than the mesh generating point,
the first moment still vanishes by construction (though not so for the other higher order odd moments).
Therefore,
we expect the gradient model to be applicable to all of the abovementioned methods.

\begin{figure*}
	\centering
	\includegraphics[trim = 10mm 10mm 10mm 10mm, clip, width=1.0\linewidth]{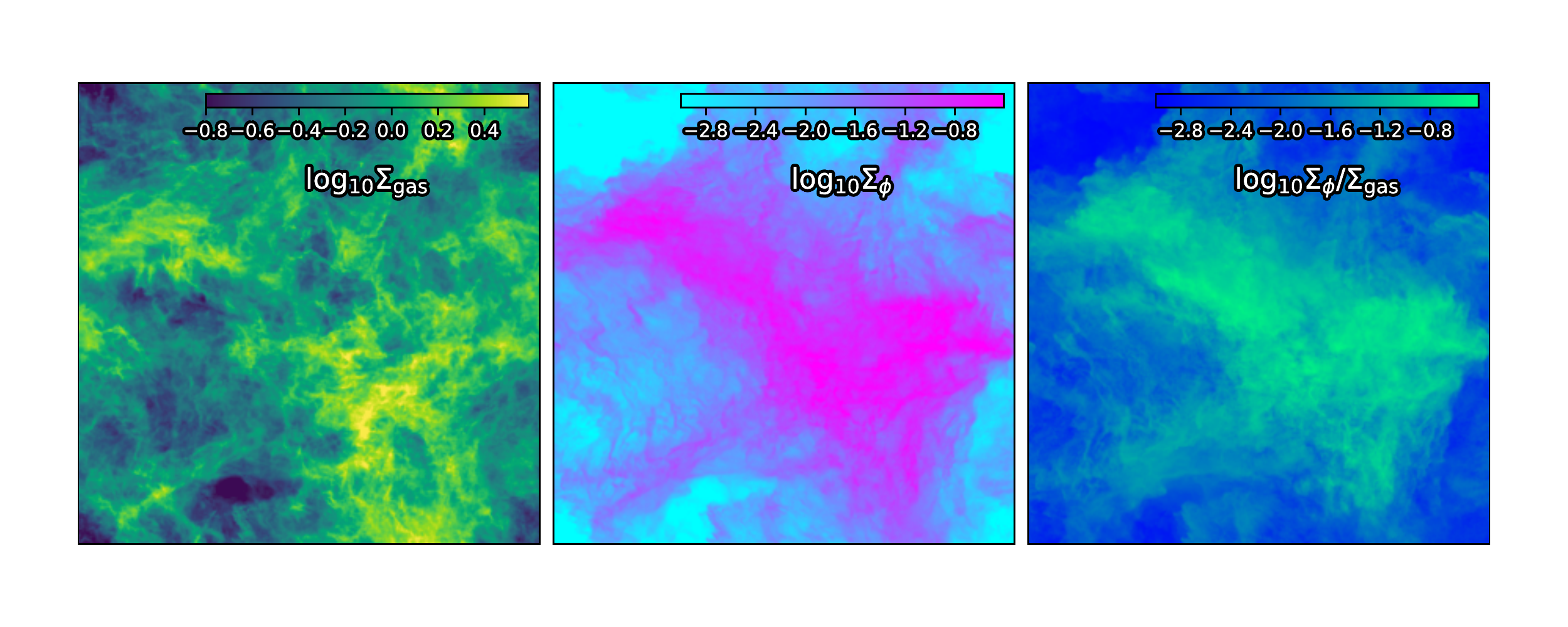}
	\caption{ Isotropic homogeneous turbulence
		with the rms Mach number $\mathcal{M} \sim 8$ (solenoidal driving) at time $t=2.2 t_{\rm eddy}$,
		where $t_{\rm eddy}$ is the eddy-turnover time.
		The three panels show maps of
		(left) gas column density $\Sigma_{\rm gas}$, 
		(middle) scalar column density $\Sigma_{\phi}$,
		and (right) $\Sigma_{\phi} / \Sigma_{\rm gas}$.
		The scalar field initially follows a radially Gaussian distribution in spherical symmetry.
		Turbulent motions deform and stretch the scalar field into a filamentary structure,
		which is very different from the diffusion picture where the field would remain spherically symmetric.
	}
	\label{fig:box0222}
\end{figure*}

\section{Validation}\label{sec:results}
\subsection{Numerical Simulations}

To test the SGS models,
we conduct high-resolution hydrodynamical simulations
and post-process the results to obtain the filtered fields
which gives us the true SGS terms
(Eqs. \ref{eq:true_tau}, \ref{eq:true_qe} and \ref{eq:true_qphi})
as well as
the SGS terms in the ED model 
(Eqs. \ref{eq:eddy_tau}, \ref{eq:eddy_qe} and \ref{eq:eddy_qphi})
and the gradient model
(Eqs. \ref{eq:grad_tau}, \ref{eq:grad_qe} and \ref{eq:grad_qphi}).
This is known as the ``\textit{a priori} test'' in the literature of turbulence modeling.
The simulations are done with the publicly available {\sc Gizmo} code \citep{2015MNRAS.450...53H},
a multi-method solver based on the meshless Godunov method \citep{2011MNRAS.414..129G} 
and built on the TreeSPH code {\sc Gadget} \citep{2005MNRAS.364.1105S}.
We adopt the meshless finite mass (MFM) solver \citep{2015MNRAS.450...53H}
which is a Lagrangian solver where cells are advected with the flows. 
The smoothing length, which corresponds to the local spatial resolution,
is determined such that it contains a fixed number of nearest cells/particles $N_{\rm ngb} = 32$.
The filter width is chosen to be around four times the MFM smoothing length
by enclosing the nearest $N_f = 4^3 N_{\rm ngb} = 2048$ particles.
We have verified that our results are insensitive to the choice of the filter size (see Appendix \ref{app:filtersize}).
The discretized coarse-grained field at $\ve{x}_a$ is calculated following the formulation from SPH:
\begin{equation}
\ol{f}_a = \sum_b \frac{m_b}{\rho_b} f_bW(\ve{x}_a - \ve{x}_b, h_a)
\end{equation}
where $f_b \equiv f(\ve{x}_b)$ and the summation is over the nearest $N_f$ particles.
The gradients are calculated with the least-square approach as in the {\sc Gizmo} code \citep{2015MNRAS.450...53H},
which is exact for a linear function irrespective of particle configuration.

\subsection{Linear Regression}
We use linear regression to assess the models and determine the best-fit coefficients.
we define the chi-squares as
\begin{align}
\chi_\tau^2   &= \sum_{b=1}^{N} |\hat{\tau}   -   \tau|^2_b = \sum_{b=1}^{N}\sum_{i=1}^3\sum_{j=1}^3(a_\tau\breve{\tau}_{ij} - \tau_{ij})^2_b, \label{eq:taufit}\\
\chi_q^2 	  &= \sum_{b=1}^{N} |\hat{\ve{q}}      -      \ve{q}|^2_b = \sum_{b=1}^{N}\sum_{i=1}^3 (a_q\breve{q}_i - q_i)^2_b, \\
\chi_\theta^2 &= \sum_{b=1}^{N} |\hat{\ve{\theta}} - \ve{\theta}|^2_b = \sum_{b=1}^{N}\sum_{i=1}^3 (a_\theta\breve{\theta}_i - \theta_i)^2_b.
\end{align}
where $\breve{\tau}_{ij}\equiv\hat{\tau}_{ij}/a_\tau$, $\breve{q}_i\equiv\hat{q}_i/a_q$,
$\breve{\theta}_i\equiv\hat{\theta}_i/a_\theta$ and 
$N$ is the number of sampled particles.
The definitions are independent of translation and rotation as they involve only the magnitudes of vectors or tensors.
We do not consider the covariance between particles and that between tensor components.\footnote{
Note that $\tau$ is symmetric and has only six degrees of freedom. 
Therefore, if the $\chi^2$ were calculated using the covariance matrix, it would be equivalent to summing over only the six independent tensor components in Eq. \ref{eq:taufit}. However, we choose not to do so as such a definition is not rotationally invariant.}
As such, the absolute values of chi-squares may be underestimated.
However, as our goal is to compare the two SGS models, the absolute values of chi-squares are unimportant.
The best-fit coefficients can be obtained analytically by minimizing the chi-squares,  
$\partial{\chi_\tau^2} / \partial{a_\tau} 
= \partial{\chi_q^2} / \partial{a_q} 
= \partial{\chi_\theta^2} / \partial{a_\theta} = 0$, 
which leads to\footnote{Note that here $i$ and $j$ are summed over as implied by Einstein notation.}
\begin{align}
\hat{a}_\tau   &= \frac{\sum_{b=1}^{N} (\breve{\tau}_{ij} \tau_{ij})_b}{\sum_{b=1}^{N} (\breve{\tau}_{ij} \breve{\tau_{ij}})_b}, \\
\hat{a}_q 	 &= \frac{\sum_{b=1}^{N} (\breve{q}_i q_i)_b}            {\sum_{b=1}^{N} (\breve{q}_i \breve{q_i})_b}, \\
\hat{a}_\theta &= \frac{\sum_{b=1}^{N} (\breve{\theta}_i \theta_i)_b}  {\sum_{b=1}^{N} (\breve{\theta}_i \breve{\theta_i})_b}.
\end{align}
The overhat notation $\hat{a}$ is to emphasize that they are the best-fit coefficients.
As the ED model only models $\m{\tau}^*$ rather than the full $\m{\tau}$,
we adopt $\chi_\tau^2   = \sum_{b=1}^{N} |\hat{\m{\tau}}^* -   \m{\tau}^*|^2_b$ for the ED model.
In order for the chi-squares to be small,
not only the magnitude but also the orientation and shape of the modeled SGS terms have to be close to the true SGS terms.

\begin{figure*}
	\centering
	\includegraphics[width=0.99\linewidth]{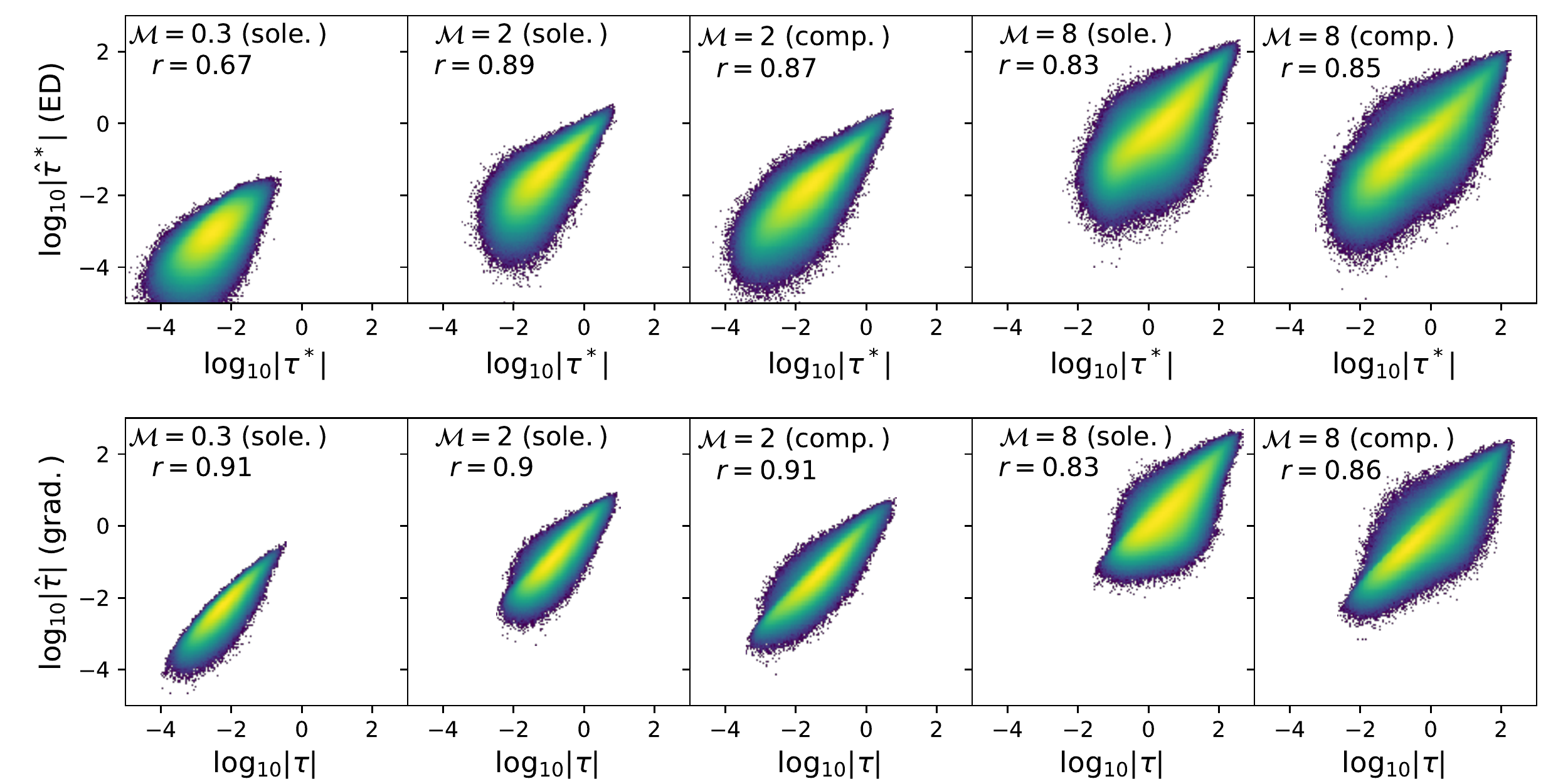}
	\caption{2D histograms of the true SGS stress tensor vs. modeled SGS stress tensor in the turbulent box.
		Panels from left to right show runs of $\mathcal{M} \sim 0.3$ (solenoidal), $\mathcal{M} \sim 2$ (solenoidal), $\mathcal{M} \sim 2$ (compressive), $\mathcal{M} \sim 8$ (solenoidal) and $\mathcal{M} \sim 8$ (compressive).
		The upper panels are for the ED model while the lower panels are for the gradient model.
		The correlation coefficient $r(|\m{\tau}|,|\hat{\m{\tau}}|)$ is shown on the upper left of each panel.		
	}
	\label{fig:figtaumagheatmaptot}
\end{figure*}

\subsection{Homogeneous Isotropic Turbulence}

We perform simulations of homogeneous isotropic turbulence assuming an isothermal equation of state.
The cubic box is initially static and uniform with density $\rho_0 = 1$ and specific thermal energy $u_0 = 1$,
which is set up with a Cartesian grid with $512^3$ equal-mass gas particles.
The box size is $L = 1$ and the boundary conditions are periodic.
The gas is continuously stirred by an external driving force using the implementation from \citet{2012MNRAS.423.2558B}.
The external driving operates in two different modes: solenoidal and compressive,
which is realized by force decomposition in the Fourier space{\footnote{It is also possible to have a mixed mode with a prescribed fraction of solenoidal and compressive driving. However, we shall only study the two extreme cases in this work.}}.
The system reaches a steady state with an rms Mach number $\mathcal{M}$.
We study five cases:
$\mathcal{M} \sim 0.3$ (solenoidal),
$\mathcal{M} \sim 2$ (solenoidal and compressive),
and $\mathcal{M} \sim 8$ (solenoidal and compressive).
Each simulation runs for 6 $t_{\rm eddy}$ where $t_{\rm eddy} = L / \mathcal{M}$ is the eddy-turnover time.
The initial passive scalar is in spherical symmetry and follows a radial distribution $\phi_0(R) = \exp(-25R^2)$ where $R$ is the radial coordinate.
As time evolves,
the scalar field will eventually become uniformly distributed and therefore does not provide further information on turbulent mixing 
(i.e., $\ve{q}$ becomes essentially zero everywhere).
As such,
we reset the scalar field back to its initial distribution at a fixed time interval of $\Delta t = t_{\rm eddy}$.
Our isothermal assumption implies that $\ve{\theta} = 0$ everywhere and therefore we cannot test the model for SGS heat flux in this setup.

\begin{table}\centering
	\begin{tabular}{| c | c | c | c | c | c |}
		\hline\hline
		Model        & $\mathcal{M}$   & $\hat{a}_{\tau}$ & $\chi^2(\tau)$ & $\hat{a}_q$ & $\chi^2(q)$ \\
		\hline
		ED 			  &         0.3 (sole.)     &      0.0398    &    4.96e-5     &    0.0642    &    2.48e-6\\
		gradient      &         0.3 (sole.)     &      0.158     &    2.28e-5     &    0.137     &    8.56e-7\\
		ED 			  &           2 (sole.)     &      0.0916    &    5.40e-2     &    0.0611    &    1.08e-4\\
		gradient      &           2 (sole.)     &      0.168     &    2.53e-2     &    0.159     &    2.56e-5\\
		ED 			  &           2 (comp.)     &      0.105     &    8.24e-3     &    0.0675    &    4.28e-5\\
		gradient      &           2 (comp.)     &      0.176     &    4.77e-3     &    0.148     &    1.21e-5\\
		ED 			  &           8 (sole.)     &      0.143     &    96.4        &    0.122     &    3.62e-3\\
		gradient      &           8 (sole.)     &      0.211     &    49.3        &    0.229     &    1.45e-3\\
		ED 			  &           8 (comp.)     &      0.147     &    10.7        &    0.132     &    9.32e-4\\
		gradient      &           8 (comp.)     &      0.204     &    5.77        &    0.220     &    3.68e-4\\
		\hline\hline
	\end{tabular}
	\caption{
		Best-fit coefficients and the validation $\chi^2$ of all five simulations of the turbulent box,
		each of which is fitted by both the ED model and the gradient model.
		The gradient model gives a better fit than the ED model in all cases.}
	\label{tab:coeffs}
\end{table}

In Fig. \ref{fig:box0222}, we show the run 
of $\mathcal{M} \sim 8$ (solenoidal) at time $t=2.2 t_{\rm eddy}$.
The three panels show maps of
(left) gas column density $\Sigma_{\rm gas}$, 
(middle) scalar column density $\Sigma_{\phi}$,
and (right) $\Sigma_{\phi} / \Sigma_{\rm gas}$.
Turbulence deforms and stretches the fluid into filamentary structures.
Serving as a tracer and advecting with the fluid,
the scalar field demonstrates that the mixing process is highly anisotropic.
This is very different from the isotropic diffusion picture which would preserve the spherical symmetry during the mixing process,
and is consistent with \citet{2017MNRAS.467.2421C} who demonstrated that the diffusion picture only works in a statistical sense.

\begin{figure*}
	\centering
	\includegraphics[trim = 5mm 0mm 5mm 0mm, clip, width=0.99\linewidth]{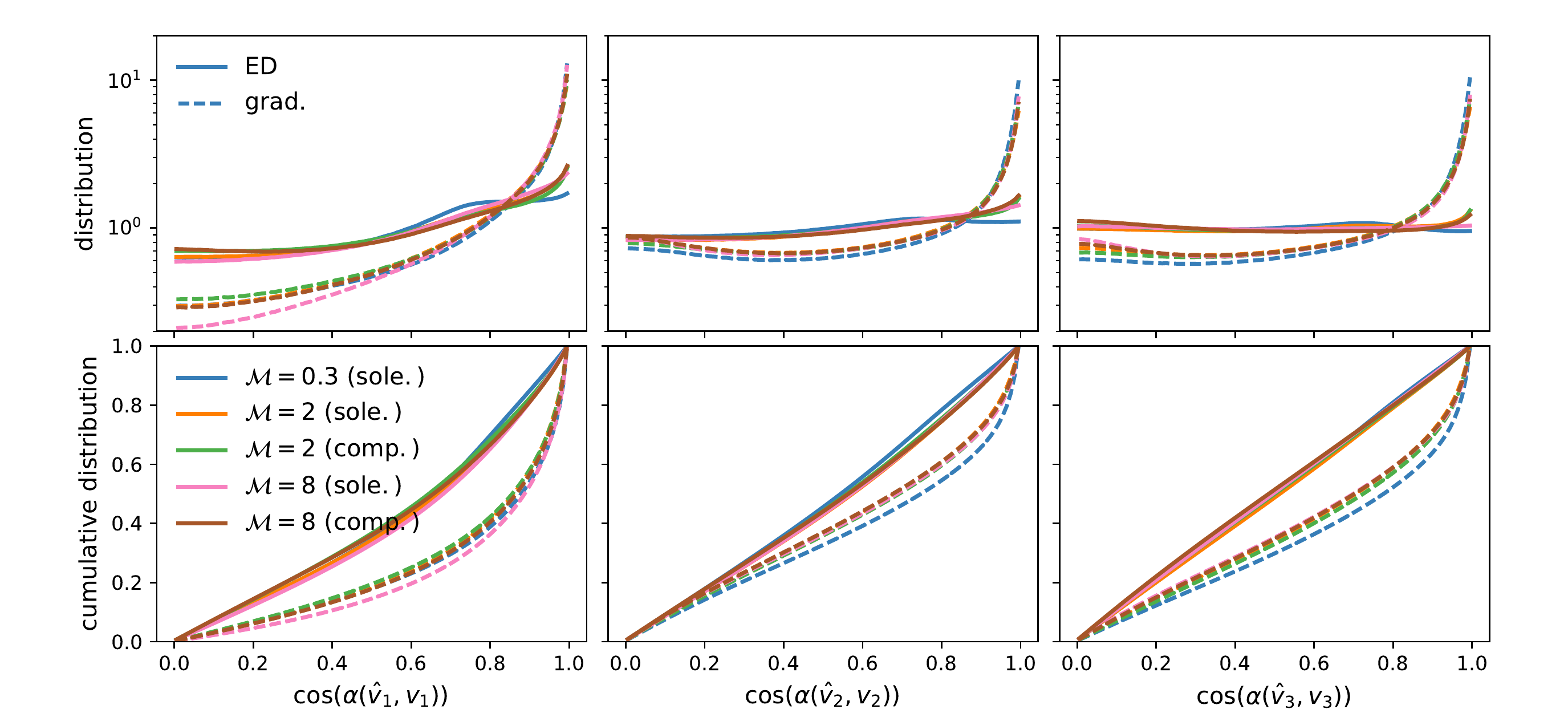}
	\caption{Distribution of
		$|\cos(\alpha( \hat{\ve{ v}} , \ve{v}))|$ in the turbulent box,
		where $\alpha( \hat{\ve{ v}} , \ve{v})$ is the angle between the true and modeled eigenvectors.
		Panels from left to right are for $\ve{ v}_1$, $\ve{ v}_2$ and $\ve{ v}_3$ 
		(corresponding to eigenvalues in a descending order), respectively.
		The normalized distribution is shown in the upper row while the corresponding cumulative distribution in the lower row.
		Here and throughout this work, we will use solid lines for the ED model and dashed lines for the gradient model.
		All three eigenvectors in the gradient model 
		are significantly better aligned with the true eigenvectors than are those in the ED model.
		The alignment is remarkably robust to variations of both Mach number and driving mechanism.
	}
	\label{fig:figtaucosinetot}
\end{figure*}

We randomly sample $512^3/32$ particles per snapshot and take 80 snapshots with a time interval of 0.05 $t_{\rm eddy}$ 
for a global fit with the number of samples $N\sim 3.3\times10^8$.
For robustness,
we use 80\% of the sample as the training set to find the best-fit coefficients,
and use the other 20\% as the validation set to calculate the $\chi^2$ for model comparison.
In practice, however,
the $\chi^2$ in both sets are very similar.
Table \ref{tab:coeffs} shows the best-fit coefficients and the validation $\chi^2$ of all the five runs,
each of which is fitted by both the ED model and the gradient model.
In all cases,
we find that the gradient model gives a better fit than the ED model, 
with about a factor of two to four times smaller $\chi^2$ in the former case.
The coefficients $\hat{a}_{\tau}$ and $\hat{a}_q$ for the gradient model fall in the range of [0.16,0.21] and [0.14,0.23], 
while those for the ED model fall in the range of [0.04,0.15] and [0.06,0.13], respectively.
In general,
the coefficients increase slightly with the rms Mach number (more significant in the ED model),
but they are insensitive to the driving mechanism (i.e., compressive or solenoidal).
The turbulent Schmidt number, defined as $\hat{a}_{\tau} / \hat{a}_q$, falls in the range of [0.92, 1.15] for the gradient model and [0.62, 1.55] for the ED model.

\subsubsection{SGS Stress Tensor}

Fig. \ref{fig:figtaumagheatmaptot} shows 2D histograms of the true SGS stress tensor vs. modeled SGS stress tensor.
Panels from left to right show runs of $\mathcal{M} \sim 0.3$ (solenoidal), $\mathcal{M} \sim 2$ (solenoidal) and $\mathcal{M} \sim 2$ (compressive), $\mathcal{M} \sim 8$ (solenoidal) and $\mathcal{M} \sim 8$ (compressive).
The upper panels are for the ED model while the lower panels are for the gradient model.
The correlation coefficient $r$ is shown in the upper left of each panel.
	The correlation coefficient is calculated in linear space rather than in log-space, viz.,
	\begin{equation}
	r(x,y) \equiv \frac{\sum_{b=1}^N (x_b - \Braket{x}) (y_b - \Braket{y})}{\sqrt{\sum_{b=1}^N \big(x_b - \Braket{x}\big)^2 \sum_{b=1}^N \big(y_b - \Braket{y} \big)^2}},
	\end{equation}
	where $\Braket{x}$ and $\Braket{y}$ are the means of $x$ and $y$, respectively.
Note that the ED model only predicts the anisotropic part of the SGS stress tensor (see Eq. \ref{eq:eddy_tau}).
Therefore, we show $|\m{\tau}^*|$ vs. $|\hat{\m{\tau}^*}|$ in the top row.
For the $\mathcal{M} \sim 0.3$ case,
the gradient model shows a much higher correlation coefficient ($r\sim 0.91$) than that in the ED model ($r\sim 0.67$).
which may explain the lower $\chi^2$ in the former case.
However,
for the $\mathcal{M} \sim 8$ cases,
both models correlate equally well with the true SGS tensor ($r \sim 0.85$),
but the gradient model still gives a lower $\chi^2$.

\begin{figure*}
	\centering
	\includegraphics[width=0.99\linewidth]{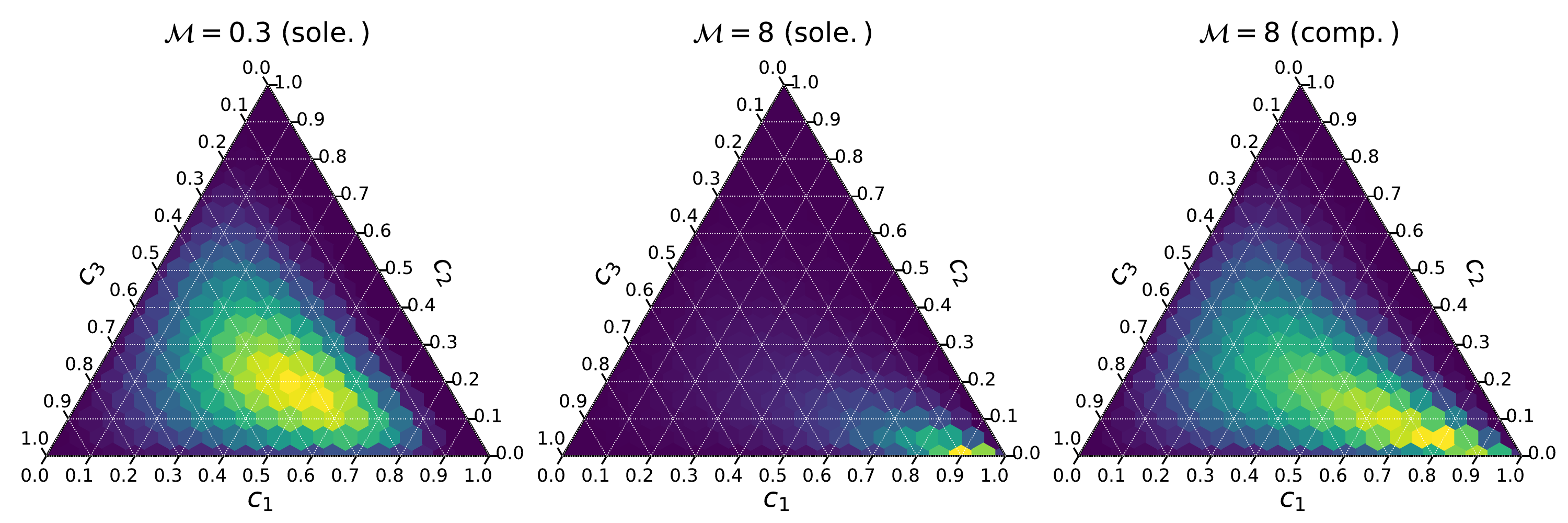}
	\caption{
			Distribution of the true shape parameters on a ternary diagram for runs
			$\mathcal{M} \sim 0.3$ (solenoidal), $\mathcal{M} \sim 8$ (solenoidal) and $\mathcal{M} \sim 8$ (compressive) from left to right.
			Only two of the three parameters are independent as $c_1 + c_2 + c_3 = 1$.
			As the Mach number increases,
			the ellipsoid becomes increasingly prolate, in particular, in the solenoidal run.
	}
	\label{fig:ternarytrue}
\end{figure*}

\begin{figure}
	\centering
	\includegraphics[width=0.99\linewidth]{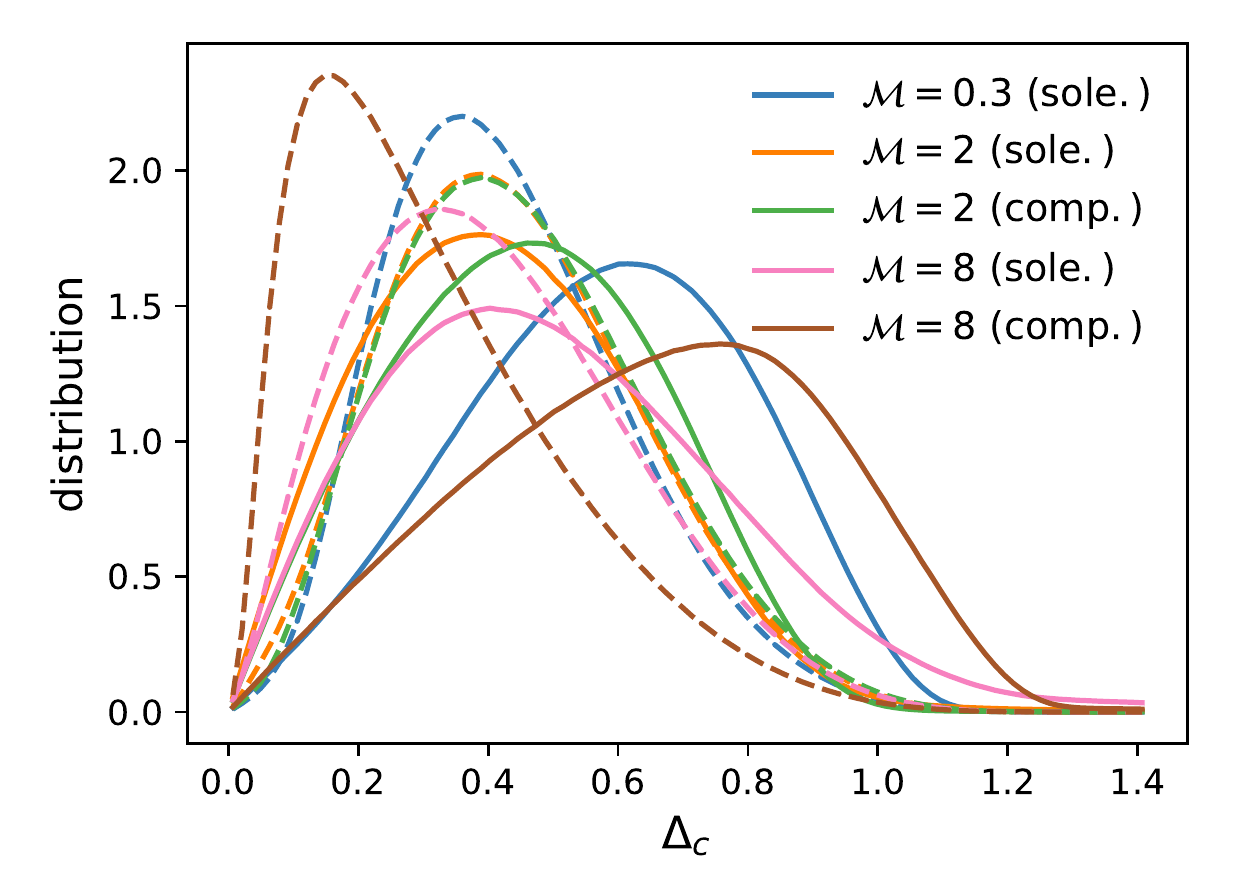}
	\caption{Normalized distribution of shape error $\Delta_c$ for both models in the turbulent box.
		The gradient model (dashed lines) makes a better prediction on the shape (i.e., smaller $\Delta_c$) than the ED model (solid lines) .
	}
	\label{fig:figtaushapedistternarytot}
\end{figure}

\begin{figure*}
	\centering
	\includegraphics[width=0.99\linewidth]{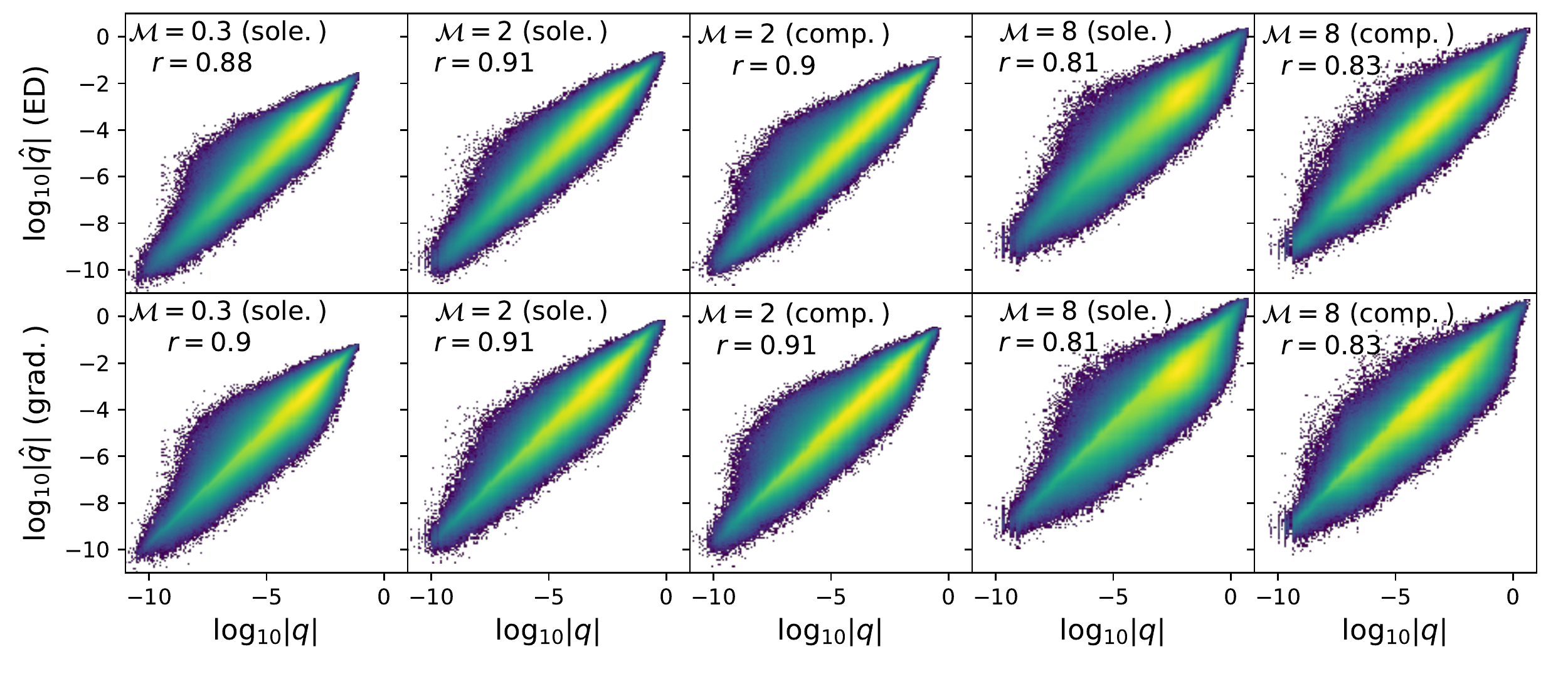}
	\caption{2D histograms of the true SGS scalar flux vs. the modeled SGS scalar flux in the turbulent box.
		Panels from left to right show runs of $\mathcal{M} \sim 0.3$ (solenoidal), $\mathcal{M} \sim 2$ (solenoidal) and $\mathcal{M} \sim 2$ (compressive), $\mathcal{M} \sim 8$ (solenoidal) and $\mathcal{M} \sim 8$ (compressive).
		The upper panels are for the ED model while the lower panels are for the gradient model.
		The correlation coefficient $r(|\ve{q}|,|\hat{\ve{q}}|)$ is shown in the upper left of each panel.
	}
	\label{fig:figqmagheatmaptot}
\end{figure*}

In fact,
the superior fit of the gradient model is due to its better prediction on the orientation and shape of $\m{\tau}$,
which we define as follows.
As $\m{\tau}$ is by construction symmetric (i.e., $\tau_{ij} = \tau_{ji}$),
its eigenvectors must be orthogonal with each other.
Moreover,
as our kernel function is strictly positive in real space,
$\m{\tau}$ must be positive semi-definite (i.e., all eigenvalues are non-negative; \citealp{1994JFM...278..351V}).
Such a tensor forms an ellipsoid $x_i \tau_{ij} x_j = 1$.
Let the eigenvectors of $\m{\tau}$ be $\ve{v}_1$, $\ve{v}_2$ and $\ve{v}_3$
and the corresponding eigenvalues be $\lambda_1$, $\lambda_2$ and $\lambda_3$ such that 
$\lambda_1 \geq\lambda_2 \geq \lambda_3 \geq 0$,
we can define the following shape parameters:
\begin{align}
c_1 = \frac{\lambda_1 - \lambda_2}{\lambda_1 + \lambda_2 + \lambda_3},\\
c_2 = \frac{2 (\lambda_2 - \lambda_3)}{\lambda_1 + \lambda_2 + \lambda_3},\\
c_3 = \frac{3 \lambda_3}{\lambda_1 + \lambda_2 + \lambda_3}.
\end{align}

The eigenvectors determine the orientations of the ellipsoid's semiaxes,
while $c_1$, $c_2$ and $c_3$ indicate, respectively, how prolate, oblate and spherical the ellipsoid is.
Note that $c_1 + c_2 + c_3 = 1$ and so only two of the three parameters are independent.
The six degrees of freedom,
one for the magnitude, three for the orientation and two for the shape,
uniquely define the tensor $\tau_{ij}$.
Moreover,
these properties are coordinate independent,
which makes them more robust in contrast to the matrix elements.

While the gradient model predicts the tensor $\m{\tau}$,
the ED model only predicts the anisotropic part $\m{\tau}^*$
which is traceless and is not positive semi-definite.
In order to compare the tensor structure in a similar way,
we add the true isotropic part of $\m{\tau}$ to the modeled $\hat{\m{\tau}}^{\rm * ED}$, viz.,
\begin{equation}
\hat{\tau}_{ij}^{\rm ED} = \hat{\tau}^{\rm * ED}_{ij} + \frac{\tau_{kk}}{3} \delta_{ij}.
\end{equation}
This represents the best-case scenario assuming that the isotropic part, which requires a separate modeling, is perfectly modeled.

In Fig. \ref{fig:figtaucosinetot},
we show the distribution of
$|\cos(\alpha( \hat{\ve{ v}} , \ve{v}))|$,
where $\alpha( \hat{\ve{ v}} , \ve{v})$ is the angle between the true and modeled eigenvectors.
Panels from left to right are for $\ve{ v}_1$, $\ve{ v}_2$ and $\ve{ v}_3$, respectively.
The normalized distribution is shown in the upper row while the corresponding cumulative distribution in the lower row.
Here and throughout this work, we will use solid lines for the ED model and dashed lines for the gradient model.
All three eigenvectors in the gradient model 
are significantly better aligned with the true eigenvectors than those in the ED model are,
where the distribution of the former sharply peaks at $|\cos(\alpha( \hat{\ve{ v}} , \ve{v}))| \sim 1$ while that of the latter is almost flat.
The first eigenvectors are better aligned than the second and third eigenvectors in both models.
This is likely due to deviation of the true and modeled eigenvalues that would lead to a different ordering
and therefore the identities of the second and third eigenvectors are swapped.
The alignment is remarkably robust to variations of both Mach number and driving mechanism.

We now turn to the shape of $\m{\tau}$, 
which can be represented as a point in the $c_1  c_2  c_3$ parameter space lying on the plane of $c_1 + c_2 + c_3 = 1$.
Fig. \ref{fig:ternarytrue}, from left to right, shows the distribution of the true shape parameters  
on a ternary diagram for runs
$\mathcal{M} \sim 0.3$ (solenoidal), $\mathcal{M} \sim 8$ (solenoidal) and $\mathcal{M} \sim 8$ (compressive).
The corners on the right, top and left represent the asymptotic prolate, oblate and spherical ellipsoids, respectively.
As the Mach number increases,
the ellipsoid becomes increasingly prolate, in particular, in the solenoidal run.
Namely,
the SGS tensor has one large eigenvalue and two much smaller eigenvalues of similar sizes (as $c_2 \sim 0$).

The shape error can be quantified as
\begin{equation}
\Delta_c \equiv \Bigg( \sum_{i=1}^{3}(c_i - \hat{c}_i )^2 \Bigg)^{1/2},
\end{equation}
which is the distance between the two points representing the true and modeled shapes.
In Fig. \ref{fig:figtaushapedistternarytot},
we show the normalized distribution of $\Delta_c$ for both models.
The gradient model makes a better prediction on shape with a smaller $\Delta_c$ than the ED model in all cases.
However,
unlike the orientation,
the shape is sensitive to both the Mach number and the driving mechanism without any clear trend.

\subsubsection{SGS Scalar Flux}

Fig. \ref{fig:figqmagheatmaptot} shows 2D histograms of the true SGS scalar flux vs. the modeled SGS scalar flux.
Panels from left to right show runs of $\mathcal{M} \sim 0.3$ (solenoidal), $\mathcal{M} \sim 2$ (solenoidal) and $\mathcal{M} \sim 2$ (compressive), $\mathcal{M} \sim 8$ (solenoidal) and $\mathcal{M} \sim 8$ (compressive).
The upper panels are for the ED model while the lower panels are for the gradient model.
The correlation coefficient $r$ is shown on the upper left of each panel.
Both models predict the magnitude of $\ve{q}$ equally well:
the correlation coefficient is as high as $r\sim 0.9$ at $\mathcal{M} \sim 0.3$ and $\mathcal{M} \sim 2$,
dropping to $r\sim 0.8$ at $\mathcal{M} \sim 8$.
This can be understood as shocks in highly supersonic regimes leading to more discontinuities where Taylor expansion is expected to break down.
Despite the equally high correlation seen in both models,
$\chi^2$ in the gradient model is about three times smaller than that in the ED model (see Table \ref{tab:coeffs}),
indicating that the discrepancy is due to orientation.

\begin{figure}
	\centering
	\includegraphics[trim = 5mm 0mm 0mm 0mm, clip, width=0.99\linewidth]{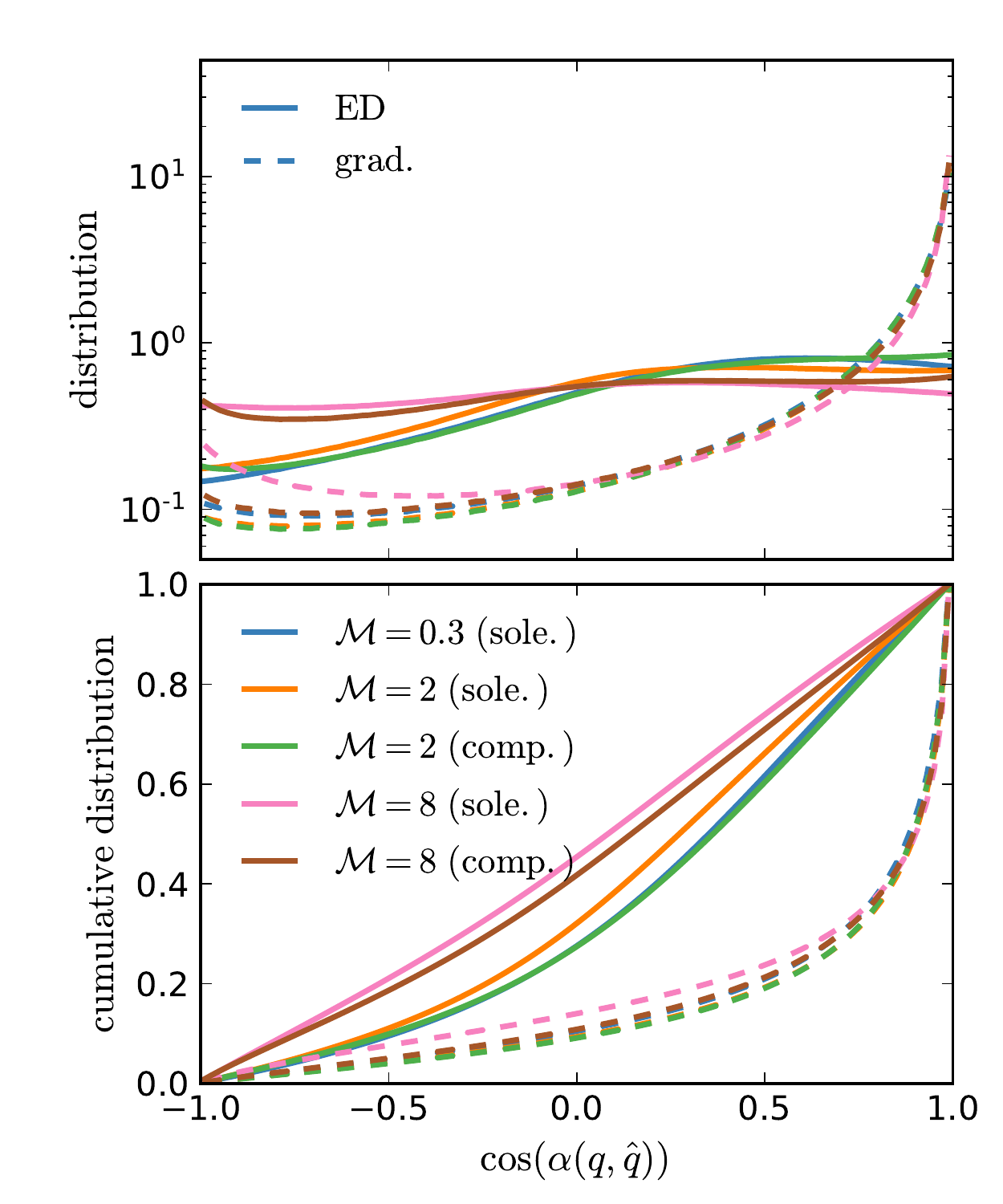}
	\caption{Distribution of
		$\cos(\alpha( \hat{\ve{ q}} , \ve{q}))$,
		where $\alpha( \hat{\ve{ q}} , \ve{q})$ is the angle between the true and modeled SGS scalar flux vectors in the turbulent box.
		The normalized distribution is shown in the upper row while the corresponding cumulative distribution is shown in the lower row.
		The gradient model predicts the orientation of $\ve{q}$ remarkably well,
		while the ED model provides little information (as much as a random vector) on the true direction of $\ve{q}$.
		Isotropic diffusion is not a good approximation of the instantaneous turbulent mixing.
	}
	\label{fig:figqcosinetot}
\end{figure}

In Fig. \ref{fig:figqcosinetot},
we show distribution of
$\cos(\alpha( \hat{\ve{ q}} , \ve{q}))$,
where $\alpha( \hat{\ve{ q}} , \ve{q})$ is the angle between the true and modeled SGS scalar flux vectors.
The normalized distribution is shown in the upper row while the corresponding cumulative distribution is shown in the lower row.
Evidently,
in all cases,
the gradient model gives a much better prediction on the orientation of $\ve{q}$ than the ED model,
which leads to a better fit.
The orientation is remarkably insensitive to both the Mach number and the driving mechanism,
especially for the gradient model.
The ED model shows an almost uniform distribution,
which means that the negative scalar gradient,
$-\partial{\wt{\phi}} / \partial{x_i}$, 
provides little information  (as much as a random vector) on the true direction of $\ve{q}$.
This demonstrates quantitatively what we alluded to earlier (see Fig. \ref{fig:box0222}) that 
isotropic diffusion is not a good approximation of the instantaneous turbulent mixing.

The better alignment between $\hat{\ve{q}}$ and $\ve{q}$ in the gradient model explains its superior fit.
It also explains why the best-fit coefficients in the ED model are smaller and show more variations than those in the gradient model:
the misalignment forces smaller best-fit coefficients in order to minimize $\chi^2$ 
as the least squares are found by a projection of the true SGS vectors onto the modeled vectors.

\begin{figure*}
	\centering
	\includegraphics[trim = 20mm 10mm 10mm 0mm, clip, width=0.99\linewidth]{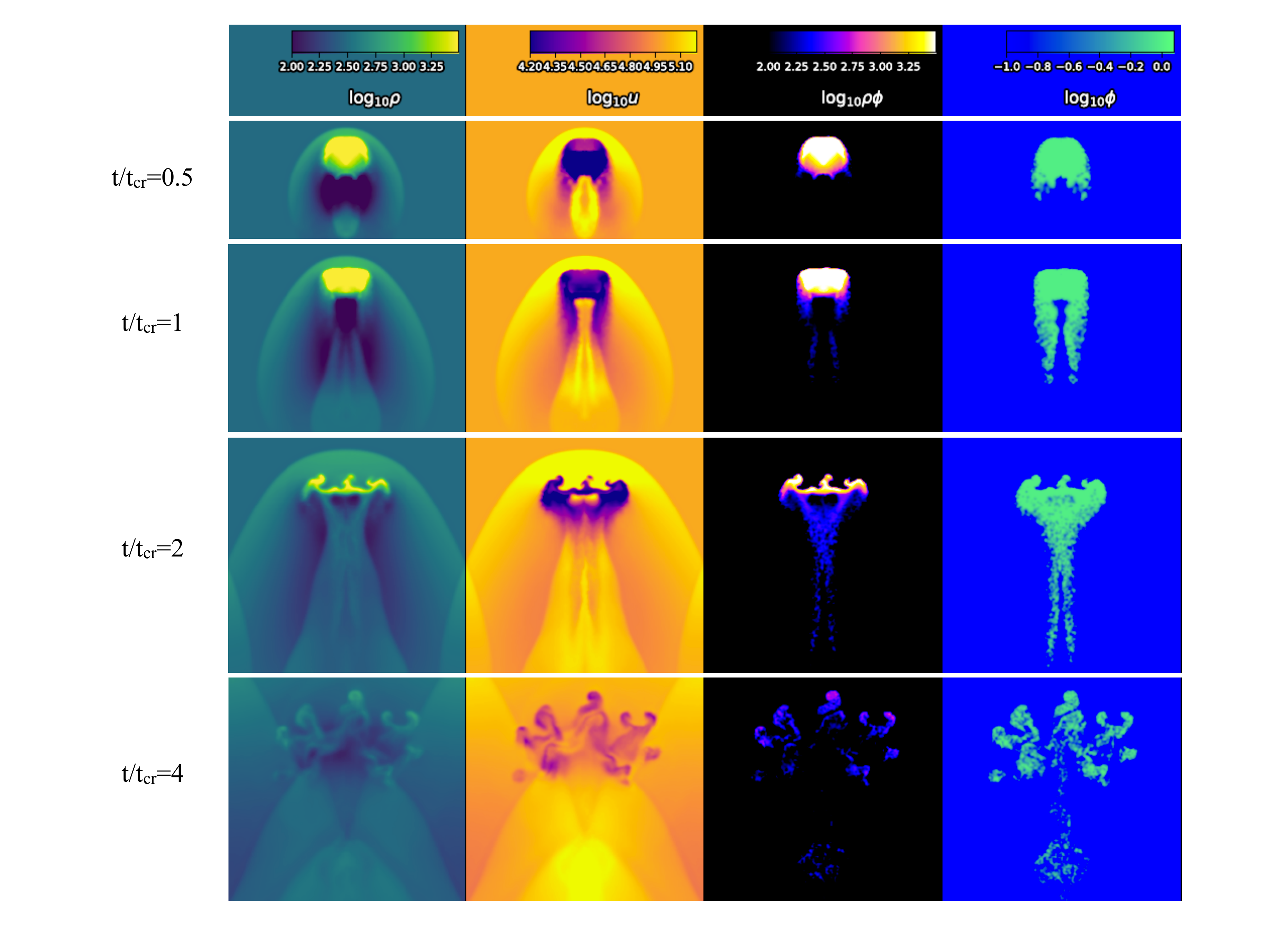}
	\caption{
		Time evolution of the wind tunnel simulation at $t/t_{\rm cr} = 0.5, 1, 2$ and 4 from top to bottom.
		Panels from left to right show slices (through the box center) of density, specific thermal energy, scalar density and scalar, respectively.
		As expected, the cloud is destroyed on a timescale of $t_{\rm cr}$ due to turbulent fluid instabilities and the two phases gradually mix with each other.
	}
	\label{fig:blobdownward}
\end{figure*}


\begin{figure}
	\centering
	\includegraphics[trim = 0mm 5mm 0mm 0mm, clip, width=0.9\linewidth]{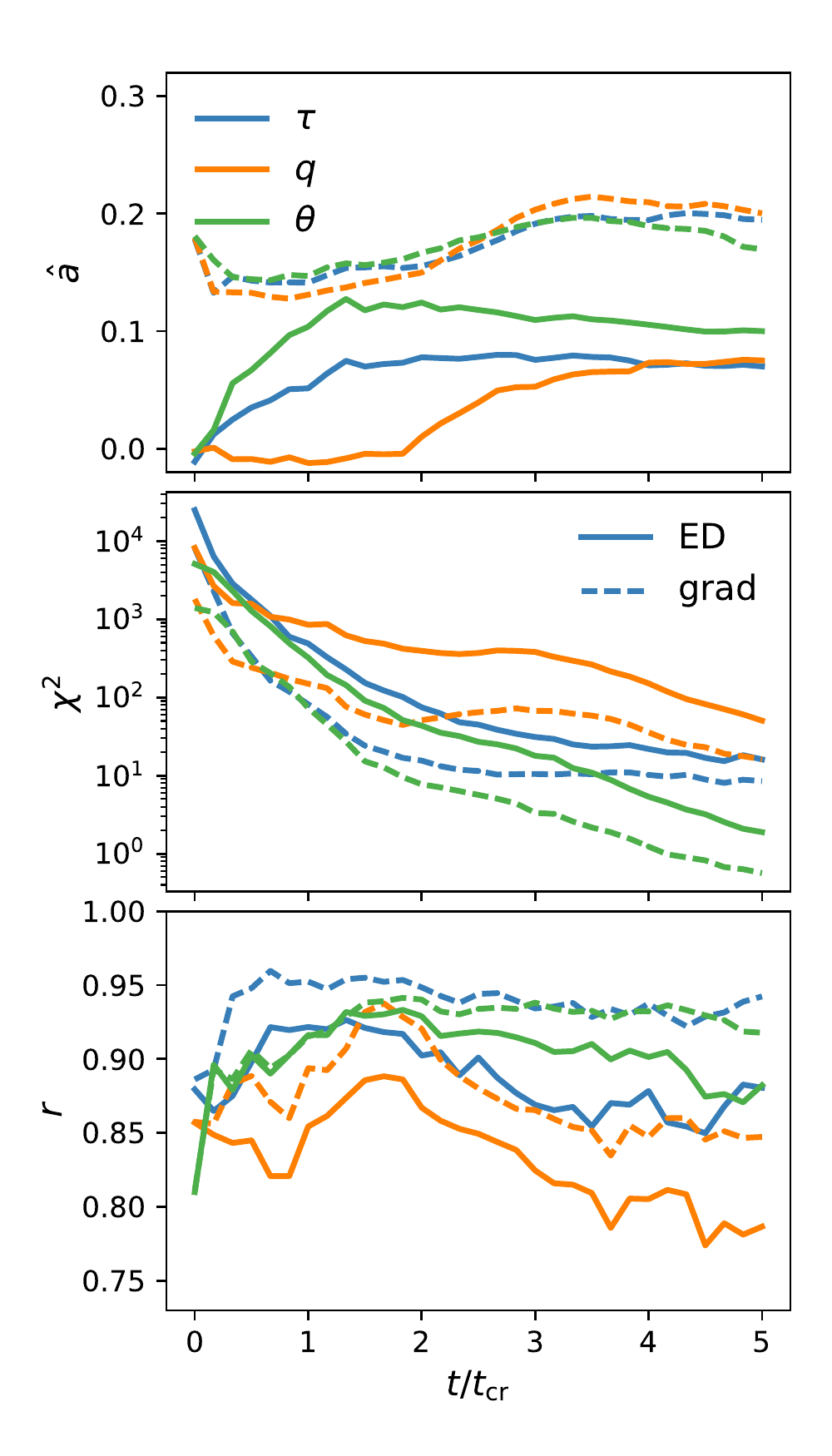}
	\caption{
		Time evolution of the best-fit coefficients (top), the corresponding $\chi^2$ (middle),
		and the correlation coefficients between the magnitudes of the true and modeled SGS terms (bottom)
		in the wind tunnel simulation.
		The gradient model provides a better fit than the ED model throughout the entire simulation.
		Strong correlations are observed in both models for all the three SGS terms,
		though the gradient model performs slightly better than the ED model.
	}
	\label{fig:figblobcorrvstime}
\end{figure}

\begin{figure*}
	\centering
	\includegraphics[trim = 10mm 0mm 10mm 0mm, clip, width=0.9\linewidth]{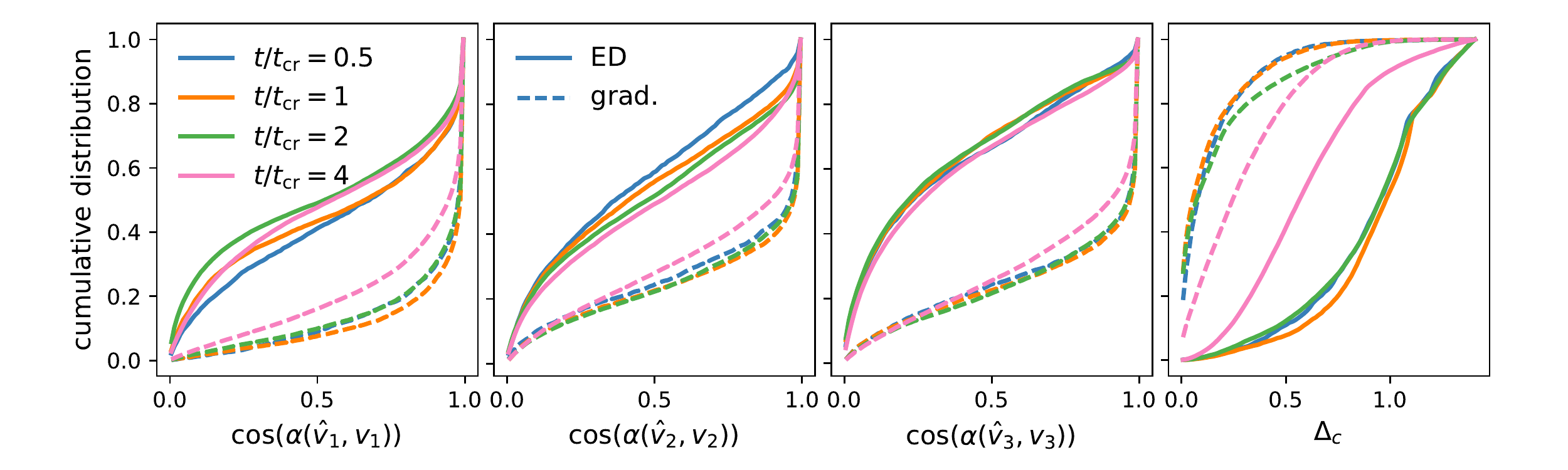}
	\caption{
		First three panels: 
		cumulative distribution of
		$|\cos(\alpha( \hat{\ve{ v}} , \ve{v}))|$, 
		where $\alpha( \hat{\ve{ v}} , \ve{v})$ is the angle between the true and model eigenvectors,
		in the wind tunnel at $t/t_{\rm cr} = 0.5, 1, 2$ and 4.
		Panels from left to right are for $\ve{ v}_1$, $\ve{ v}_2$ and $\ve{ v}_3$ (corresponding to eigenvalues in a descending order), respectively.
		Rightmost panel: cumulative distribution of shape error $\Delta_c$.
		The gradient model makes a much better prediction on both the orientation and the shape of $\m{\tau}$ than the ED model does throughout the simulation.
	}
	\label{fig:figblobtaucosine}
\end{figure*}


\begin{figure}
	\centering
	\includegraphics[width=0.9\linewidth]{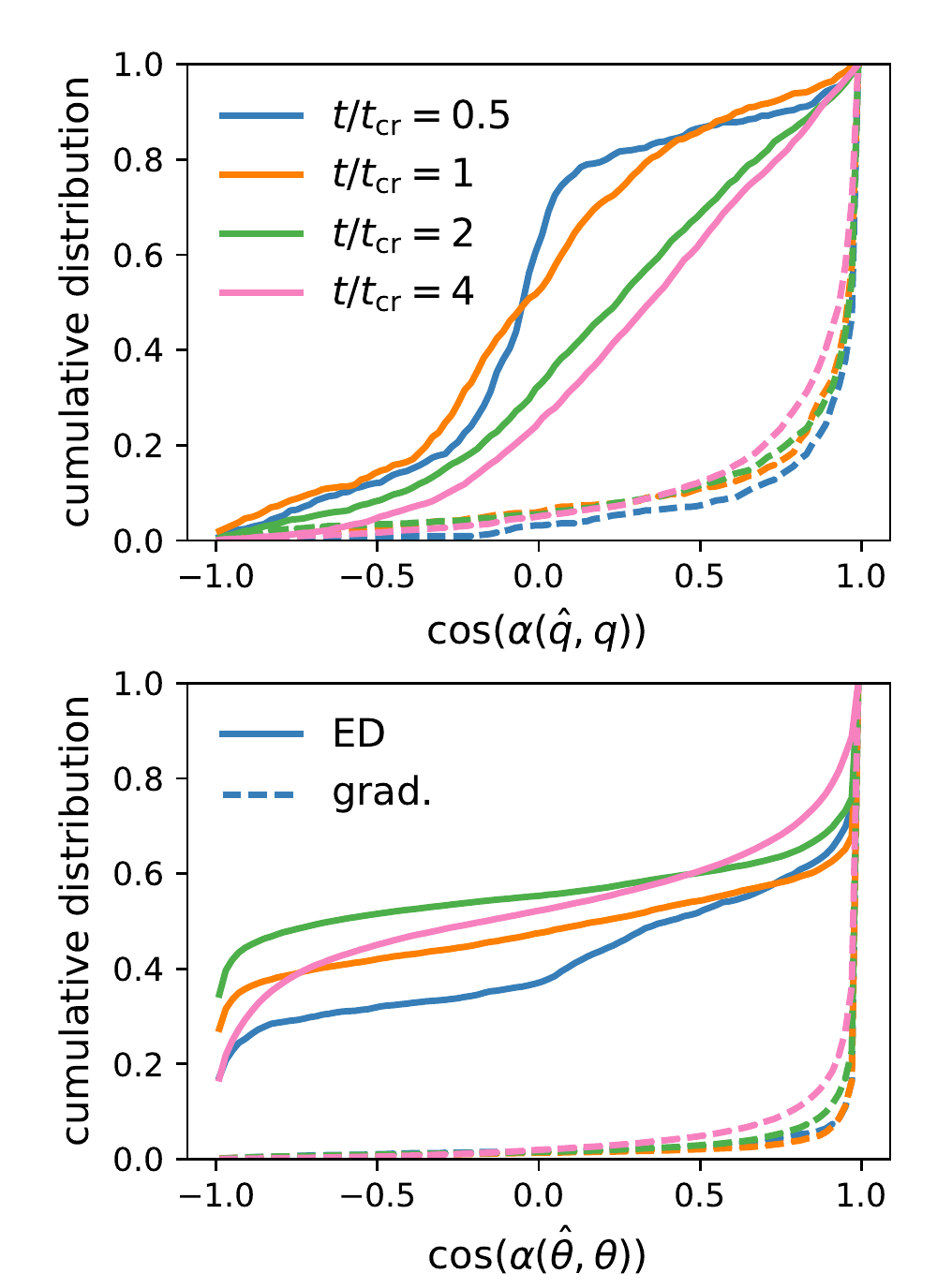}
	\caption{
		Upper: cumulative distribution of $\cos(\alpha( \hat{\ve{ q}} , \ve{q}))$
		in the wind tunnel at $t/t_{\rm cr} = 0.5, 1, 2$ and 4,
		where $\alpha( \hat{\ve{ q}} , \ve{q})$ is the angle between the true and model SGS scalar flux vectors.
		Lower: same as the upper panel but for the SGS heat flux ${\rm cos}(\alpha( \hat{\ve{\theta}} , \ve{\theta}))$.
		The gradient model makes a much better prediction on the orientation of $\ve{q}$ and $\ve{\theta}$ than the ED model does throughout the simulation.
	}
	\label{fig:figblobqcosine}
\end{figure}

\subsection{Wind Tunnel}\label{sec:blob}

The idealized turbulent box simulations allow us to test the SGS models with well-controlled Mach number and driving mechanism.
In most realistic applications, however,
turbulent flows are rarely in statistical equilibrium.
In this section,
we analyze the classical wind tunnel simulation,
a transient setup frequently encountered in astrophysics 
where a cold and dense cloud traveling in a hot and diffuse background gas 
(e.g., \citealp{1994ApJ...420..213K, 1994ApJ...433..757M, 2007MNRAS.380..963A, 2008MNRAS.388.1079I}).
A cloud with radius $R_c = 200$, density $\rho_c = 3\times 10^{-7}$ and specific thermal energy $u_c = 10^4$ is moving along the elongated axis at $v_c = 1000$
through the background gas with density $\rho_b = 3\times 10^{-8}$ and specific thermal energy $u_b = 10^5$.
The box size is $[L_x, L_y, L_z] = [2000, 2000, 6000]$ and the boundary conditions are periodic.
The inter-particle length of the background is $\Delta_b = 14$ while that of the cloud is $\Delta_b = 6.5$.
Namely,
the cloud diameter is resolved by $\sim 60$ particles.
The equation of state is adiabatic and the radiative cooling is not included.
Due to turbulent fluid instabilities,
the cloud will gradually dissolve and mix with the background gas
on a cloud-crushing timescale $t_{\rm cr} = 2 (R_c / v_c) (\rho_c / \rho_b)^{1/2}$ = 1.26.
The simulation runs for $5t_{\rm cr}$.

Fig. \ref{fig:blobdownward} shows the time evolution of the system at $t/t_{\rm cr} = 0.5, 1, 2$ and 4 from top to bottom.
Panels from left to right show slices (through the box center) of density, specific thermal energy, scalar density and scalar, respectively.
As expected, the cloud is destroyed on a timescale of $t_{\rm cr}$ due to turbulent fluid instabilities and the two phases gradually mix with each other.

As the system is obviously not in a steady state,
we fit each snapshot separately instead of stacking all snapshots together for a global fit as we did for the turbulent box.
In Fig. \ref{fig:figblobcorrvstime},
we show the time evolution of the best-fit coefficients (top), the corresponding $\chi^2$ (middle),
and the correlation coefficients between the magnitudes of the true and modeled SGS terms (bottom)
in the wind tunnel simulation.
As time evolves,
$\chi^2$ decreases not because of the model becomes more accurate with time,
but because the SGS terms decrease when the two phases gradually mix.
The gradient model provides a better fit than the ED model throughout the entire simulation.
Coefficients for the ED model show more significant variations with time than those for the gradient model.
In fact,
the ED model even requires negative coefficients occasionally, especially during the early stage.
This is again due to the misalignment of the modeled SGS terms with respect to the true SGS terms.
On the other hand,
the coefficients in the gradient model for all three SGS terms are remarkably close to each other,
where the turbulent Schmidt number ($\hat{a}_{\tau} / \hat{a}_q$) and
the turbulent Prandtl number ($\hat{a}_{\tau} / \hat{a}_{\theta}$) are in the range of [0.92, 1.12] and [0.93, 1.15], respectively.
This can happen when most of the modeled vectors misaligned with the true vectors by almost 180$\textdegree$.
We find strong correlations between the true and modeled magnitudes in both models for all three SGS terms,
though the gradient model performs slightly better than the ED model.

In the first three panels of Fig. \ref{fig:figblobtaucosine},
we show 
the cumulative distribution of
$|\cos(\alpha( \hat{\ve{ v}} , \ve{v}))|$, 
where $\alpha( \hat{\ve{ v}} , \ve{v})$ is the angle between the true and model eigenvectors,
in the wind tunnel at $t/t_{\rm cr} = 0.5, 1, 2$ and 4.
Panels from left to right are for $\ve{ v}_1$, $\ve{ v}_2$ and $\ve{ v}_3$ (corresponding to eigenvalues in a descending order), respectively.
The distribution shows little time evolution in both models.
The gradient model provides an excellent prediction on the orientations, even better than that in the turbulent box.
At $t/t_{\rm cr}=4$,
there is a slight deterioration of accuracy presumably because $\m{\tau}$ approaches zero as the mixing completes.
However,
the distribution is still sharply peaked around one.
The same trend can be observed in the shape error.
shown in the rightmost panel of Fig. \ref{fig:figblobtaucosine}.
The majority of particles fall in the region of $\Delta_c < 0.1$ in the gradient model,
meaning that their shape is in excellent agreement with the true $\m{\tau}$-tensor.
On the other hand,
the ED model performs poorly in both orientation and shape,
similar to what we find in the turbulent box.

We now turn to $\ve{q}$ and $\ve{\theta}$.
In the upper panel of Fig. \ref{fig:figblobqcosine},
we show the cumulative distribution of $\cos(\alpha( \hat{\ve{ q}} , \ve{q}))$
in the wind tunnel at $t/t_{\rm cr} = 0.5, 1, 2$ and 4,
where $\alpha( \hat{\ve{ q}} , \ve{q})$ is the angle between the true and modeled SGS scalar flux.
The lower panel of Fig. \ref{fig:figblobqcosine} shows the same but for the SGS heat flux ${\rm cos}(\alpha( \hat{\ve{\theta}} , \ve{\theta}))$.
Again,
the gradient model provides excellent predictions on the orientations of $\ve{q}$ and $\ve{\theta}$,
even better than in the turbulent box.
In addition,
the distribution shows little time evolution in the gradient model
except for a slight shift toward lower values with time,
which can be understood in a similar way as discussed above.
On the other hand,
the ED model provides poor predictions on the orientation for both $\ve{q}$ and $\ve{\theta}$.
This is especially so for the SGS scalar flux as $\ve{\hat{q}}$ appears to be randomly oriented with respect to $\ve{q}$.
On the other hand,
for the SGS heat flux,
there is a large fraction ($\sim$40\%) of cells in the range of ${\rm cos}(\alpha( \hat{\ve{\theta}} , \ve{\theta})) > 0.9$ (i.e. highly aligned),
but there is also a similar fraction of cells in the range of ${\rm cos}(\alpha( \hat{\ve{\theta}} , \ve{\theta})) < -0.9$ (i.e. completely opposite direction),
indicating a significant back scattering.
This qualitative difference between $\ve{q}$ and $\ve{\theta}$ is intriguing 
as it indicates that the mixing of specific thermal energy does not behave like that of passive scalars, 
presumably due to shock heating and adiabatic expansion.

\section{Summary and Conclusions}\label{sec:sum}

We have conducted high-resolution numerical simulations and coarse-grained the results to validate turbulent SGS models of momentum, energy and passive scalars.
Two models are investigated:
the ED model 
(Eqs. \ref{eq:eddy_tau}, \ref{eq:eddy_qe} and \ref{eq:eddy_qphi})
and the gradient model
(Eqs. \ref{eq:grad_tau}, \ref{eq:grad_qe} and \ref{eq:grad_qphi}).
For the isothermal turbulent box in statistical equilibrium,
the gradient model consistently provides a better fit than the ED model for both $\hat{\m{\tau}}$ and $\hat{\ve{q}}$ 
in subsonic ($\mathcal{M}\sim0.3$), transonic ($\mathcal{M}\sim 2$) and supersonic ($\mathcal{M}\sim8$) regimes
(Table \ref{tab:coeffs}).
In terms of magnitude,
the modeled SGS terms correlate equally well with the true SGS terms in both models (Figs. \ref{fig:figtaumagheatmaptot} and \ref{fig:figqmagheatmaptot}).
However,
the gradient model makes a much better prediction on the orientation and shape of the SGS terms (Figs. \ref{fig:figtaucosinetot}, \ref{fig:figtaushapedistternarytot} and \ref{fig:figqcosinetot}),
which explains its superior fit.
The correlation of orientation is remarkably insensitive to both the Mach number and the driving mechanism,
while the correlation of shape is sensitive to both.
The strong misalignment between the modeled and true SGS terms in the ED model leads to smaller and more variations of its best-fit coefficients.

A similar trend is observed in the wind tunnel simulation which is transient and highly time-dependent.
The gradient model gives a better fit throughout the simulation (Fig. \ref{fig:figblobcorrvstime}).
The modeled magnitude shows comparable correlations with the true magnitude,
with the gradient model performs slightly better.
However, 
in terms of orientation and shape,
the gradient model provides a much better prediction compared to the ED model 
(Figs. \ref{fig:figblobtaucosine} and \ref{fig:figblobqcosine}).



When applying the ED model for $\m{\tau}$,
one must bear in mind that it only models the anisotropic part $\m{\tau}^*$ instead of the complete tensor.
The isotropic part would require a separate modeling, 
which in principle can be done by adding another transport equation for the SGS turbulent kinetic energy $k$ (e.g. \citealp{2011A&A...528A.106S}).
Ignoring the isotropic part (e.g. \citealp{2019MNRAS.483.3810R}) is effectively assuming $k=0$ which is not recommended.
On the other hand,
the gradient model has the advantage that it models the complete tensor $\m{\tau}$ 
and thus there is no need for a separate model for $k$.

The true SGS fluxes rarely align with the negative gradient of the transport variables.
Instantaneous turbulent transport is highly anisotropic and hence isotropic diffusion becomes a poor approximation (Fig. \ref{fig:box0222}).
This makes sense because the diffusion approximation only applies if the mean free path of the random walk is much smaller than the length scale of interest,
which is not the case in turbulence.
Furthermore,
in realistic astrophysical applications,
one often encounters flows that are transient and out of statistical equilibrium where turbulence is anisotropic and not fully developed.
In such cases,
the ED model is not expected to be a good approximation even in a statistical sense.
Computationally,
switching from the ED model to the gradient model adds little extra cost
as the velocity gradient is already required in the former case when constructing the velocity shear.
Therefore,
we advocate using the gradient model with $\hat{a}_{\tau}$ in the range of [0.16,0.21],
and the turbulent Schmidt number ($\hat{a}_{\tau} / \hat{a}_q$) and Prandtl number ($\hat{a}_{\tau} / \hat{a}_{\theta}$) both
in the range of [0.92,1.15] (Table \ref{tab:coeffs} and Fig. \ref{fig:figblobcorrvstime}).


In practice,
numerical diffusion may overwhelm the effects of the SGS terms,
making them pointless to be included.
This is especially so in astrophysical applications where one typically adopts upwind Riemann solvers 
in order to deal with highly compressible flows and shock capturing.
Therefore,
numerical diffusion is in general more severe than in terrestrial applications where less dissipative schemes are often used (e.g. central difference or spectral methods). 
Furthermore, 
numerical diffusion is expected to be both solver- and problem-dependent.
This uncertainty can be ameliorated by enlarging the scale on which the SGS terms operate (e.g. on a few resolution elements)
until it dominates over numerical diffusion.
However,
this is at the price of the turbulent inertial range that can still be followed in simulations and a compromise has to be made.
A systematic study on the effect of numerical diffusion relative to the explicit SGS models is required.
One exception is the scalar transport in Lagrangian simulations where the numerical scalar diffusion is by construction zero and thus the SGS model is always recommended.

\section*{Acknowledgments}
We thank the anonymous referee for their valuable and constructive comments that improved our manuscript.
We thank Volker Springel and Phil Hopkins for making the {\sc Gadget} and {\sc Gizmo} codes publicly available.
We use {\sc pygad}\footnote{https://bitbucket.org/broett/pygad} for visualization
and {\sc python-ternary}\footnote{https://github.com/marcharper/python-ternary}
for generating the ternary diagrams.
The Center for Computational Astrophysics is supported by the Simons Foundation.
C.Y.H. acknowledges support from the DFG via German-Israel Project Cooperation grant STE1869/2-1 GE625/17-1.


\bibliography{literatur}{}
\bibliographystyle{aasjournal}

\begin{figure*}
	\centering
	\includegraphics[width=0.8\linewidth]{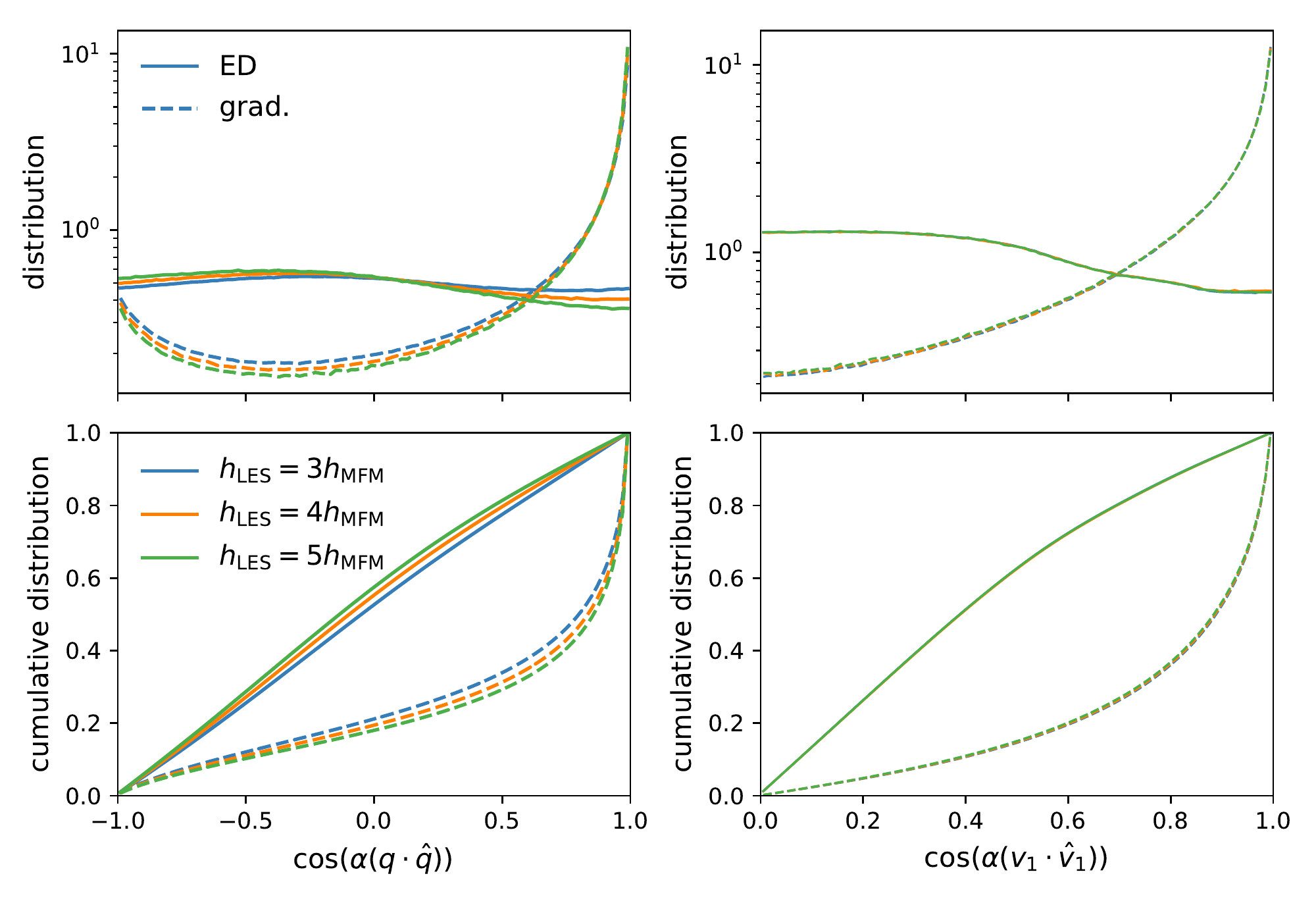}
	\caption{
			The distribution (top) and cumulative distribution (bottom) of 
			$\cos(\alpha( \hat{q} , q))$ (left) and 
			$|\cos(\alpha( \hat{v}_1 , v_1))|$ (the first eigenvector, right) for the $\mathcal{M} \sim 8$ solenoidal case
			in the time period of $[3t_{\rm eddy}, 4t_{\rm eddy}]$,
			done with three different filter sizes, 
			$h_{\rm LES} = 3 h_{\rm MFM}$, $4 h_{\rm MFM}$ and $5 h_{\rm MFM}$.
			The results are insensitive to the choice of $h_{\rm LES}$.
	}
	\label{fig:fhsmlqtau}
\end{figure*}

\appendix

\section{Filter size}\label{app:filtersize}

We explore how sensitive our results are to the choice of the filter size in this section.
We redo a subset of the analysis for the $\mathcal{M} \sim 8$ solenoidal case in the time period of 
$[3t_{\rm eddy}, 4t_{\rm eddy}]$
with three different filter sizes, 
$h_{\rm LES} = 3 h_{\rm MFM}$, $4 h_{\rm MFM}$ and $5 h_{\rm MFM}$
where $h_{\rm MFM}$ is the cell size (smoothing length) of the simulation.
Fig. \ref{fig:fhsmlqtau} shows the distribution (top) and cumulative distribution (bottom) of 
$\cos(\alpha( \hat{q} , q))$ (left) and 
$|\cos(\alpha( \hat{v}_1 , v_1))|$ (the first eigenvector, right), respectively.
The results are insensitive to the choice of $h_{\rm LES}$,
especially for $|\cos(\alpha( \hat{v}_1 , v_1))|$.



\end{document}